# Fluorescence Molecular Tomography for Quantum Yield and Lifetime

**WENXIANG CONG AND GE WANG\***

*Biomedical Imaging Center, Department of Biomedical Engineering, Rensselaer Polytechnic Institute, Troy, NY 12180 USA*
*\*wangg6@rpi.edu*

**Abstract:** Fluorescence molecular tomography (FMT) is a promising modality for non-invasive imaging of internal fluorescence agents in biological tissues especially in small animal models, with applications in diagnosis, therapy, and drug design. In this paper, we present a new fluorescent reconstruction algorithm that combines time-resolved fluorescence imaging data with photon-counting micro-CT (PCMCT) images to estimate the quantum yield and lifetime of fluorescent markers in a mouse model. By incorporating PCMCT images, a permissible region of interest of fluorescence yield and lifetime can be roughly estimated as prior knowledge, reducing the number of unknown variables in the inverse problem and improving image reconstruction stability. Our numerical simulation results demonstrate the accuracy and stability of this method in the presence of data noise, with an average relative error of 18% in fluorescent yield and lifetime reconstruction.

## 1. Introduction

Optical molecular imaging is a widely used tool for diagnosing diseases, evaluating therapies, and developing drugs in small animal models. This imaging modality analyzes the interaction of light with biological tissues to estimate the concentration of fluorescent or bioluminescent tracers in biological issues especially a live mouse [1, 2]. Fluorescence molecular tomography (FMT) is a technique for 3D fluorescent source reconstruction from photon fluence rates collected on the surface of the mouse body for quantifying fluorescent biomarkers of diseases such as cancer in a mouse model. Time-domain excitation is typically performed with short laser pulses in conjunction with time-resolved data acquisition, providing preclinically relevant information about optical properties inside objects.

Time-resolved data directly reveal the intrinsic fluorophore lifetime of excited molecular probes. The fluorescence quantum yield, which is the ratio of the number of fluorescent photons emitted to the number of photons absorbed. This yield helps visualize the concentration and distribution of fluorophores, and the location and dynamics of gene expression and molecular interaction in tissues. The fluorescence lifetime, which changes with different quenching mechanisms related to the biological microenvironment (local pH, calcium or sodium ion concentrations, oxygen saturation, etc.), is sensitive to various pathological processes and robust with respect to experimental parameters such as fluorophore concentration and excitation intensity [3, 4]. Therefore, fluorescence lifetime imaging is more advantageous than fluorescence intensity imaging.

The fluorescence intensity on the tissue surface is related to the concentration of fluorophores, optical properties, and the tissue anatomy. Micro-CT can provide detailed anatomical structures of small animals. Moreover, the recent development of photon-counting micro-CT (PCMCT) allows for material decomposition, offering contrast enhancement and improved anatomical prior. Different kinds of tumors in a live mouse can be detected by PCMCT to delineate permissible fluorescent source regions and stabilize FMT results [5].

Photon propagation in biological tissues can be modeled using the radiative transfer equation (RTE) [6] or Monte Carlo simulation [7]. Accurate optical parameters are crucial for high-quality tomographic imaging, while inaccurate parameters would compromise the localization and quantification of molecular probes [8]. A novel method has been developed to estimate optical parameters using PCMCT [9]. Specifically, multi-energy images of biological tissues obtained from PCMCT can be segmented into different organs and tissue constituents. The optical characteristics of these organs and constituents can then be used to calculate their absorption and scattering coefficients [9].

The highly scattering nature of near-infrared light in biological tissues makes it challenging to reconstruct fluorescent yield and lifetime tomographically [10, 11]. To address this problem, gradient-based optimization with Tikhonov regularization was used in the image reconstruction [12, 13]. However, this optimization may have multiple local solutions, and the effectiveness of this method depends on the initial conditions, model approximation, measurement noise, and algorithmic design. Various regularization methods were proposed to find a meaningful solution to this inverse problem [11, 14]. Tikhonov regularization incorporates a structural prior constraint on the underlying image; for example, the $L_1$ norm and its variants yield sparse solutions [15, 16], total variation (TV) minimization promotes piecewise smoothness [17], non-convex $L_q$ (0<q<1) and log function-based regularizations often produce optical tomographic reconstruction quality better than that with the L2 regularization [18-20]. However, due to the ill-conditioned characteristic and the limitations of fluorescence measurements in small animal imaging, these reconstruction techniques often rely on the choice of regularization parameters and the number of iterations, resulting in uncertain results.

In this study, we present a new algorithm for fluorescent molecular tomography that utilizes time-resolved measurement data on the surface of the mouse body and PCMCT images to estimate the quantum yield and lifetime of fluorescent probes. In Section 2, we introduce the imaging geometry and physical models, and present algorithms for tomographic reconstruction of fluorescence yield and lifetime. In Section 3, we perform numerical experiments to evaluate the accuracy and stability of the proposed method. Finally, in Section 4 we discuss relevant issues and conclude this paper.

## 2. Methodology

### 2.1 PCMCT to identify tumor locations

In preclinical mouse studies, PCMCT scans a live mouse and acquires projection data from various projection views in multiple energy bins. Then, PCMCT images can be reconstructed in multi-energy bins from projection data and segmented to identify various features inside the mouse [5]. These images provide sufficient information about anatomical structures and chemical compositions of tissues, allowing for better estimation of optical absorption and scattering coefficients of each involved tissue than what can be done using current-integrating micro-CT [9]. With contrast-enhanced PCMCT, material decomposition can be performed based on these multi-energy micro-CT images to identify tumors [21]. Effectively, contrast agents enhance the visibility of organs and tumors [22]. This allows PCMCT to identify the tumor location and volume, greatly facilitating the image reconstruction process in the context of fluorescence molecular tomography, as illustrated in Figure 1 [22].

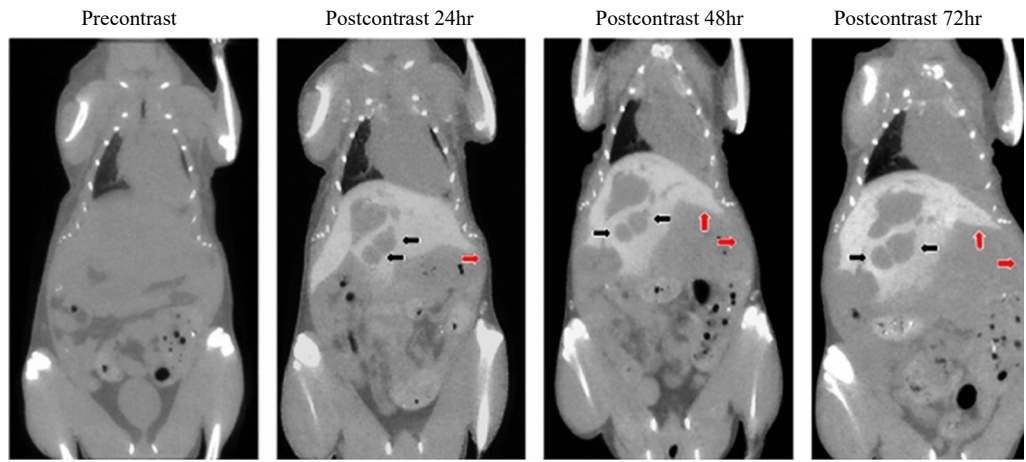

**Fig. 1.** B16F10 tumor growth in the mouse liver observed using Fenestra HDVC. The murine melanoma cells grew in the mouse for two weeks before the mouse was given a single dose of Fenestra HDVC. Using in vivo micro-CT, the quantity and volume of tumors in the liver were measured for three days. The tumors measured for volume are indicated by the black arrows and the abnormal protrusion in the abdominal cavity. The liver is shown by the red arrows. Adapted from "In vivo micro-computed tomography imaging in liver tumor study of mice using Fenestra VC and Fenestra HDVC, Scientific Reports 12, 22399 (2022)" by Ming Jia Tan, Nazarine Fernandes, Karla Chinnery Williams & Nancy Lee Ford [22].

*2.2 Process of fluorescence tomographic imaging*

To label a biological target with fluorescent molecules, fluorescent agents are injected into a small animal to generate fluorescent contrast. After a suitable time period, the fluorescent biomarkers accumulate sufficiently in targeted tissues such as tumors due to the targeting property of fluorescent agents. As a result, the concentration of the fluorescent agent in a region of interest is much higher than in the tissue background. The fluorescent biomarkers are then excited by an external pulsed laser source. Short pulses with a wavelength in the excitation range of the fluorophore illuminate the animal. These incoming photons propagate in the tissues, some of which reach the targets and excite the fluorophores inside them. Subsequently, the excited fluorophores emit fluorescent photons with a longer wavelength than the excitation wavelength. The emitted photons propagate through the tissues, some of which escape the small animal surface and are measured by the time-resolved optical detectors with appropriate filters.

In a non-contact imaging setup, a CCD camera is positioned at a distance from the mouse. To perform fluorescence imaging, an external near-infrared laser is directed toward a region of interest to excite the fluorophores in the mouse. Subsequently, fluorescence photons of longer wavelengths are generated and travel through the mouse, being then captured by time-resolved photon detectors to produce high-quality fluorescence views of the body surface of the mouse, as shown in Fig. 2. This process is sequential, starting from laser pulsing, radiation from the source to the mouse surface, excitation of internal fluorophores, emission of fluorescent photons, and propagation of the emitted photons in the mouse body to the detector. Each measurement dataset is then mapped onto the surface of the animal based on the imaging geometry, and the data is the input for image reconstruction.

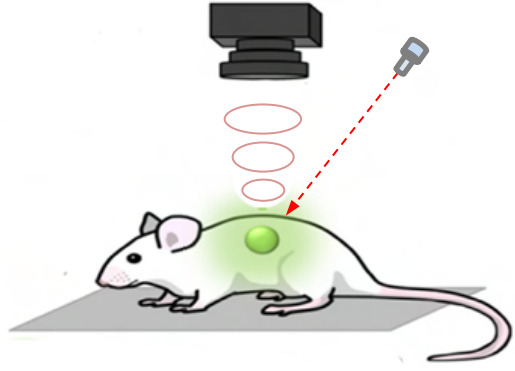

Fig. 2. Epi-illuminating fluorescence imaging using a pulse excitation light source on the same side of an optical camera for time-resolved data acquisition.

## 2.3 Model for photon transport in tissue

The physical model for fluorescence imaging can be described by coupled equations that account for the light transport processes at the excitation and emission wavelengths, respectively:

$$\frac{1}{\upsilon}\frac{\partial L_x(r,\omega,t)}{\partial t}+s\cdot\nabla L_x(r,\omega,t)+\left(\mu_a^x+\mu_a^f+\mu_s^x\right)L_x(r,\omega,t)$$
$$=\mu_s^x\int_{4\pi}p(\omega,\omega')L_x(r,s,t)d\omega'+S_x(r,\omega,t) \quad (1)$$

$$\frac{1}{\upsilon}\frac{\partial L_m(r,\omega,t)}{\partial t}+s\cdot\nabla L_m(r,\omega,t)+\left(\mu_a^m+\mu_s^m\right)L_m(r,\omega,t)$$
$$=\mu_s^m\int_{4\pi}p(\omega,\omega')L_m(r,s,t)d\omega'+S_m(r,\omega,t) \quad (2)$$

where $S_x(r,\omega,t)$ is a light source at wavelength $\lambda_x$ to excite fluorophores for fluorescence emission. The emission of the fluorescent molecules takes place at wavelengths $\lambda_m$. The fluorophore absorbs light energy of a specific wavelength and re-emits light of a longer wavelength ($\lambda_m > \lambda_x$). The absorption wavelength, energy transfer efficiency, and time before emission depend on the fluorophore structure and its chemical environment. For excited fluorophores, the fluorophore reaction can be modeled as the rate equation:

$$\frac{dR(t)}{dt}=-(1/\tau)R+\mu_a^f(r)U_x(r,t) \quad (3)$$

where $\mu_a^f$ is the fluorophore absorption coefficient, and $U_x(r,t)=\int_{4\pi}L_x(r,\omega,t)d\omega$. The solution to this equation is

$$R(t)=\mu_a^f(r)\int_0^t U_x(r,t')\exp\left(-\frac{t-t'}{\tau}\right)dt' \quad (4)$$

where the source term of the equation is the convolution of the excitation function and the fluorescence pulse response function:

$$S_m(r,t) = Q\mu_a^f(r)\int_0^t U_x(r,t')\exp\left(-\frac{t-t'}{\tau}\right)dt' \quad (5)$$

where $\tau(r)$ denotes the fluorescence lifetime, which is the average time the molecule remains in its excited state before emitting a photon and returning to the ground state, the quantum efficiency $Q$ is the ratio of fluorescent photons emitted to photons absorbed, and the fluorescence emission is directly proportional to the quantum efficiency of the fluorophore. The fluorescent yield $Q\mu_a^f(r)$ is the product of quantum efficiency and fluorophore absorption.

Using the Green function method, the time-resolved fluorescence photon fluence rate at a detector point for an impulsive excitation at a source position can be expressed as a time convolution:

$$U(r_s,r_d,t) = \int_V\left[\int_0^t G_m(r,r_d,t-t')\left(\int_0^{t'} G_x(r,r_s,t'')\exp\left(-\frac{t'-t''}{\tau(r)}\right)Q\mu_a^f(r)dt''\right)dt'\right]d^3r \quad (6)$$

where $G_x(r,r_s,t)$ is the Green function of Eq. (1), and $G_m(r,r_d,t)$ is the Green function of Eq. (2). The time-resolved fluorescence photon fluence rate $U(r_s,r_d,t)$ can be measured by the photon detectors using a fluorescence imaging system, as shown in Fig. 2.

### 2.4. Reconstruction of fluorescence yield and lifetime

**Reconstruction of lifetime**: Fluorescence lifetime is sensitive to various pathological processes and is robust with respect to experimental parameters such as fluorophore concentration and excitation intensity. We can first reconstruct the lifetime from time-resolved measurements via optimization based on Eq. (6):

$$\tau = \arg\min\sum_d \left\|\arg\max\left(U(r_s,r_d,t|\tau)\right) - \arg\max\left(m(r_s,r_d,t|\tau)\right)\right\| \quad (7)$$

where $m(r_s,r_d,t|\tau)$ is the fluorescence photon fluence rate on measurement position for time gate $t$, and $U(r_s,r_d,t|\tau)$ is the simulated fluorescence photon fluence rate using Eq. (6). Importantly, the variations in fluorescent yield do not affect lifetime reconstruction accuracy. Equation (7) can be solved using an optimization method to obtain the fluorescence lifetime from time-resolved measurements.

**Reconstruction of fluorescence yield**: By integrate both sides of Eq. (6) with respect to the time variable t, we obtain the photon fluence rate measurable on the photon detectors. Using the convolution theorem, we have the following integral equation:

$$\int_0^\infty U(r_s,r_d,t)dt = \int_V\left[\int_0^\infty G_x(r,r_s,t)dt\int_0^\infty G_m(r,r_d,t)dt\right]Q\mu_a^f(r)\tau(r)d^3r \quad (8)$$

The left-hand side of Eq. (8) is the time-independent photon fluence rate calculated via superposition along the time direction from the measured data of the time-resolved fluorescence photon fluence rate $U(r_s, r_d, t)$. From the reconstructed lifetime $\tau$, Eq. (8) can simulate time-independent photon fluence rate and perform the reconstruction of fluorescence quantum yield. Based on finite element analysis, Eq. (8) can be discretized into a matrix equation with respect to fluorescence yield $y$:

$$\Phi(\Gamma) = Ay \qquad (9)$$

where $\Gamma$ is the boundary nodal set on the finite element mesh of the animal, which corresponds to the measured photon fluence rates. $\Phi(r_d) = \int_0^\infty U(r_s, r_d, t) dt$ is calculated from the superposition of time-resolved fluorescence photon fluence rate $U(r_s, r_d, t)$ along the time direction. Based on Eq. (9), we can reconstruct fluorophore concentration from the time-resolved measurement via optimization. By incorporating PCMCT images, we first localize the tumor location and volume by identifying high contrast regions, forming permissible regions of interest for reconstruction of fluorescence yield and lifetime. By subtracting background fluorescence emitting from normal tissues [23, 24], we focus on reconstruction of the fluorescence source in the region of interest. As a result, the number of unknown variables for FMT can be greatly reduced to help overcome the underdetermined nature of the inverse problem and improve quality of reconstructed FMT images.

3. **Numerical Experiments**

A digital mouse (https://neuroimage.usc.edu/neuro/Digimouse) was used in the simulation to evaluate the proposed FMT method. The Digimouse was generated using co-registered CT and cryosection images of a 28g nude normal male mouse, and converted into a finite elemental model with 306,773 tetrahedron elements and 58,244 nodes (16,164 surface nodes and 42,080 internal nodes). The structures segmented from these data include the whole brain, external cerebrum, cerebellum, olfactory bulbs, striatum, medulla, massetter muscles, eyes, lachrymal glands, heart, lungs, liver, stomach, spleen, pancreas, adrenal glands, kidneys, testes, bladder, skeleton and skin [25]. Fluorophores (CY7) were numerically distributed in a spherical region of radius 0.8mm centered at (20.0, 49.0, 13.0) mm in the Digimouse to mimic contrast enhancement in the tumor vasculature. The closest distance from the source center to the Digimouse surface is 6.95mm. The optical properties of the Digimouse were assumed to be 0.01mm-1 for absorption coefficient and 10.0 mm-1 for scattering coefficient. The lifetime of fluorophores inside the 0.8mm-diameter sphere was set to 0.5 ns. The excitation pulse light of 710nm at the incident position (20.0, 48.8, 20.4) mm on the surface of the Digimouse was used to excite fluorophores and generate the fluorescent emission at wavelength 805nm. We developed a MATLAB code to perform the time-resolved forward simulation of the photon fluence rate measurement using Eq. (6), and built an inverse model based on Eq. (8). Fig. 3 shows the photon fluence rate of the measurable positions on the mouse surface to illustrate the accuracy of the inverse model. In the 2ns sampling step, the maximum relative difference between the photon fluence rates calculated by Eqs. (6) and (8) is about 1.65%. The tomographic reconstruction of fluorescence yield and lifetime were performed using the reconstruction approach described in Subsection 2.4. The average error of the reconstructed lifetime is about 0.02ns. Fig. 4 compares the ground truth and the reconstructed fluorescent yield without noise in terms of the photon fluence rate. Similarly, Fig. 5 compares the ground truth and the fluorescent yield reconstructed from photon fluence rate with Gaussian white noise (GWN) of 40dB. The average relative error of the reconstructed fluorescent yield is about 18.19%. The relative error of total fluorescence yield over the region of interest is 1.61%.

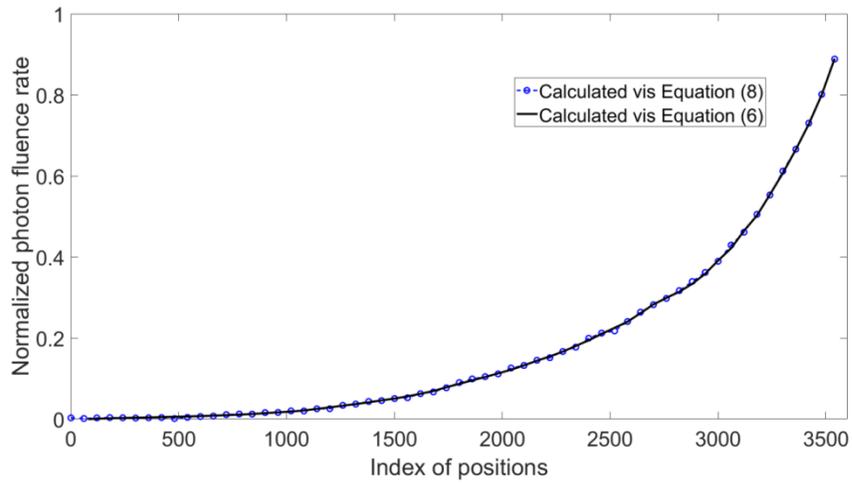

Fig.3. Photon fluence rate at various positions on the Digimouse surface.

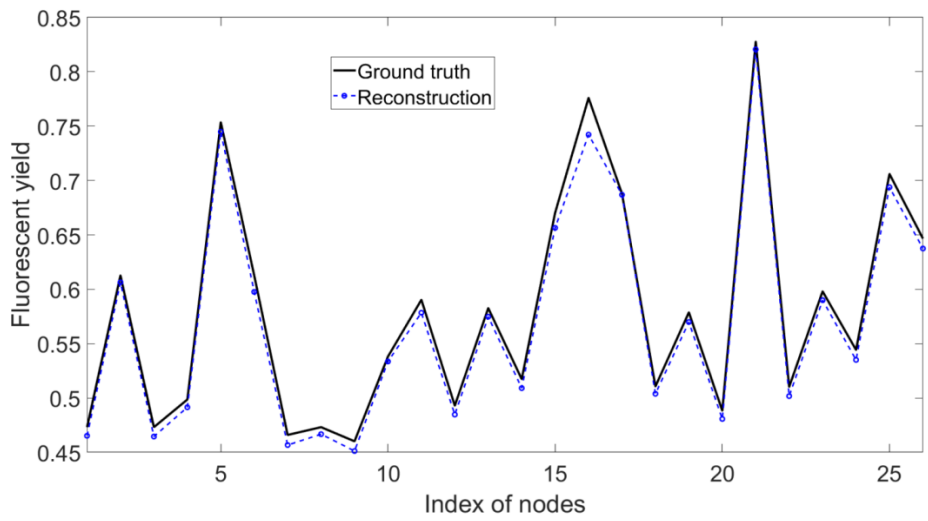

Fig.4. Fluorescent yield reconstruction of the Digimouse from photon fluence rates without data noise.

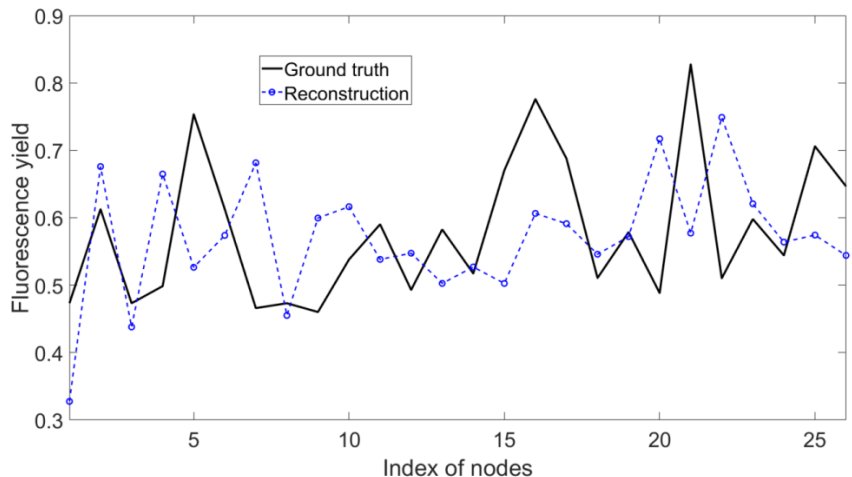

Fig.5. Fluorescent yield reconstruction of the Digimouse from noisy photon fluence rates.

## 4. Discussions and conclusion

Tomographic reconstruction of fluorescence quantum yield and lifetime is a powerful technique widely used in small animal studies to non-invasively visualize and quantify biological and pathological processes. It is well known that this inverse source problem is challenging mainly due to the high scattering nature of fluorescence photons in biological tissues. An optimization method is typically employed to solve this problem, but the ill-conditioned characteristic and photon measurement noise can result in uncertain reconstruction results, making it rather difficult to apply this technology in preclinical practice. To address this issue, we have developed a novel reconstruction algorithm to estimate the quantum yield and lifetime of fluorescent markers from time-resolved measurements on the surface of a mouse body, taking advantages of PCMCT images to identify high-contrast regions as permissible regions of interest for fluorescence yield and lifetime reconstruction. This reduces the number of unknown variables in the inverse problem and helps overcome the underdetermined nature. Our numerical experiments demonstrate the accuracy and stability of the proposed reconstruction method in the presence of data noise, achieving a reconstruction error of 0.02 ns for fluorescence lifetime and an average relative error of 18% for quantum yield reconstruction.

In the follow-up studies, we will evaluate our proposed method on live animal data and verify the results with histology. For that purpose, in vivo estimation of optical parameters will be performed based on PCMCT images and literature reference values. It is expected that some fitting or calibration techniques will be needed since optical parameters of a live mouse depend on many physiological factors and so on.  Nevertheless, the information extracted from PCMCT images should be helpful and will be evaluated in physical phantom experiments and mouse studies.

In conclusion, we have combined photon-counting x-ray imaging and fluorescence molecular imaging techniques for tomographic reconstruction of fluorescence quantum yield and lifetime in a mouse model. The technical feasibility and merits have been demonstrated in numerical simulation. This approach has a potential to significantly improve fluorescence molecular tomography and enhance our understanding of biological mechanisms of many human diseases in vivo, and should have applications in various tasks such as drug development and disease diagnosis.

**Funding.** National Institutes of Health Grants (R01CA237267).

**Acknowledgments.** The authors acknowledge the support of the National Institutes of Health Grants NIH/NIBIB R01CA237267.

**Disclosures.** Authors declare no conflicts of interest.

**Data availability.** No data were generated or analyzed in the presented research.